\documentclass[review,number,sort&compress]{elsarticle} % use 'preprint for the final verison'
\usepackage{graphicx}
\usepackage[english]{babel}
\usepackage[utf8]{inputenc}
\usepackage[T1]{fontenc}
\usepackage{siunitx}
\usepackage{subfig}
\def\ops/{\mbox{o-Ps}}
\def\degree{^{\circ}}
\journal{Nucl. Instrum. and Meth. A.}
\begin{document}
\begin{frontmatter}
\title{Trilateration-based reconstruction of ortho-positronium decays into three photons with the J-PET detector%
}%
\cortext[cor1]{Corresponding author.}
\author[WFAIS]{A.~Gajos}
\author[WFAIS]{D.~Kami\'nska}
\author[WFAIS]{E.~Czerwi\'nski\corref{cor1}}
\ead{eryk.czerwinski@uj.edu.pl}
\author[WFAIS]{D.~Alfs}
\author[WFAIS]{T.~Bednarski}
\author[WFAIS]{P.~Bia\l as}
\author[WFAIS]{B.~G\l owacz}
\author[LUBLIN]{M.~Gorgol}
\author[LUBLIN]{B.~Jasi\'nska}
\author[WFAIS,PAN]{\L.~Kap\l on}
\author[WFAIS]{G.~Korcyl} 
\author[SWIERK]{P.~Kowalski}
\author[WFAIS]{T.~Kozik}
\author[NCBJ]{W.~Krzemie\'n}
\author[WFAIS]{E.~Kubicz}
\author[WFAIS]{M.~Mohammed}
\author[WFAIS]{Sz.~Nied\'zwiecki}
\author[WFAIS]{M.~Pa\l ka}
\author[WFAIS]{M.~Pawlik-Nied\'zwiecka}
\author[SWIERK]{L.~Raczy\'nski}
\author[WFAIS]{Z.~Rudy}
\author[WFAIS]{O.~Rundel}
\author[WFAIS]{N.G.~Sharma}
\author[WFAIS]{M.~Silarski}
\author[WFAIS]{A.~S\l omski}
\author[WFAIS]{A.~Strzelecki}
\author[WFAIS,PAN]{A.~Wieczorek}
\author[SWIERK]{W.~Wi\'slicki}
\author[LUBLIN]{B.~Zgardzi\'nska}
\author[WFAIS]{M.~Zieli\'nski}
\author[WFAIS]{P.~Moskal}
\address[WFAIS]{Faculty of Physics, Astronomy and Applied Computer Science,
 Jagiellonian University, 30-348 Cracow, Poland}
\address[LUBLIN]{Department of Nuclear Methods, Institute of Physics, Maria Curie-Sk\l odowska University, 20-031, Lublin, Poland.}
\address[PAN]{Institute of Metallurgy and Materials Science of Polish Academy of Sciences, Cracow, Poland.}
\address[SWIERK]{\'Swierk Computing Centre, National Centre for Nuclear Research, 05-400 Otwock-\'Swierk, Poland}
\address[NCBJ]{High Energy Department, National Centre for Nuclear Research, 05-400 Otwock-\'Swierk, Poland}
\begin{abstract}
This work reports on a new reconstruction algorithm allowing to reconstruct the decays of ortho-positronium atoms into three photons using the places and times of photons recorded in the detector. The method is based on trilateration and allows for a simultaneous reconstruction of both location and time of the decay. Results of resolution tests of the new reconstruction in the J-PET detector based on Monte Carlo simulations are presented, which yield a spatial resolution at the level of \SI{2}{\centi\metre} (FWHM) for X and Y and at the level of \SI{1}{\centi\metre} (FWHM) for Z available with the present resolution of J-PET after application of a kinematic fit. Prospects of employment of this method for studying angular correlations of photons in decays of polarized ortho-positronia for the needs of tests of CP and CPT discrete symmetries are also discussed. The new reconstruction method allows for discrimination of background from random three-photon coincidences as well as for application of a novel method for determination of the linear polarization of ortho-positronium atoms, which is also introduced in this work.
\end{abstract}
\begin{keyword}
  \texttt{Reconstruction} \sep \texttt{Positronium} \sep \texttt{Discrete symmetry}  \sep \texttt{J-PET}
  \PACS 36.10.Dr \sep 11.30.Er \sep 24.80.+y
\end{keyword}
\end{frontmatter}
\section{Introduction}
Trilateration is a widely known technique used for determination of a position of a point known to lie simultaneously on surfaces of several spheres with given radii and centers. In a two-dimensional space, knowledge of three intersecting and non-identical circles is required in order to find a unique solution for the aforementioned point. Similarly, in a three-dimensional case, information about three spheres narrows the set of possible solutions to at maximum two points. In practical trilateration applications, additional requirements on the sought point location usually exist which allow to identify the correct solution of the possible two. 

Applications of trilateration-based localization usually determine the radii of the spheres by measuring  the times of propagation of certain types of signals exchanged between the object being localized and several reference objects whose positions are well-defined and correspond to centers of the spheres. The propagation time measurement requires both the object being localized and the reference ones to determine the moment of signal emission or arrival with respect to the same starting point. This, however, is not possible in many practical realizations where the signal arrival or emission time cannot be measured for the localized object. The classical trilateration problem has then to be extended to involve an additional unknown time and the radii of the spheres become parametrized by this variable rather than being constant. As the problem of finding an intersection of three spheres with radii defined up to variable value is underdetermined, one additional constraint is required in order to limit the set of solutions to two as in classical trilateration. The additional information can be provided by a fourth reference object, as is the case in the most widely known application, the Global Positioning System (GPS). A similar approach, however, was recently applied in particle physics for reconstruction of the $K_L$ neutral meson decays into two and three neutral pions at the KLOE experiment~\cite{mgr_gajos,gajos_appb}. In these cases photons act as the exchanged signal and the reference points are provided by the recording places of the 4 and 6 photons, respectively coming from the decays $K_L\to 2\pi^0\to 4\gamma$ and $K_L\to 3\pi^0\to 6\gamma$.

Decays with three secondary particles only, however, can be reconstructed as well in a similar manner when the additional constraint is based on the geometry of the event rather than on a fourth particle. In this paper, we present a reconstruction method based on trilateration, intended for reconstructing decays of ortho-positronium into three gamma quanta (\ops/$\to3\gamma$) for the needs of discrete symmetry tests in ortho-positronium decays.

\section{Prospects of discrete symmetry studies with the J-PET detector}
Although the CP and CPT discrete symmetries are thoroughly tested in a large variety of phenomena (see e.g.~\cite{Babusci:2013gda}), there have been only a few experiments investigating their conservation in the leptonic sector~\cite{Arbic:1988pv,Skalsey:1991vt,Conti:1993,vetter:2003,Yamazaki:2009hp}. Ortho-positronium~(\ops/), the triplet bound state of electron and positron, was pointed out as a purely leptonic system sensitive to symmetry violation effects~\cite{Bernreuther:1981ah} and several angular correlations observable in the ortho-positronium decays into three photons have been defined~\cite{Bernreuther:1988tt}. These correlations use momenta of the photons produced in the decay ordered by energy $\mathrm{|\vec{k}_{1}|>|\vec{k}_{2}|>|\vec{k}_{3}|}$ and the spin $\mathrm{\vec{S}}$ of the positronium. The two latest experiments following this scheme searched for non-zero expectation values of the CP-odd correlation $\mathrm{(\hat{S}\cdot\hat{k}_{1})(\hat{S}\cdot\hat{k}_{1}\times \hat{k}_{2})}$~\cite{Yamazaki:2009hp} and $\mathrm{(\hat{S}\cdot\hat{k}_{1}\times \hat{k}_{2})}$ sensitive to CPT violation~\cite{vetter:2003} in decays of spin-polarized ortho-positronia and both have limited the symmetry violation with a precision between~\num{e-2} and~\num{e-3}.

The J-PET device is a novel detector based on long (\SI{50}{\centi\metre}) strips of fast plastic scintillator arranged axially in a multi-layer barrel~\cite{Moskal:2013sxa,patent_jpet_strip,patent_jpet_matrix}. While \mbox{J-PET} was developed with Positron Emission Tomography in mind, it is capable of recording photons from ortho-positronium decays, thus allowing for tests of both CP and CPT symmetries with the aforementioned angular photon correlations.
Detailed description of potential of the J-PET detector for studies of discrete symmetries in decays of positronium atom can be found in Ref.~\cite{PawelActaB}.
The design of J-PET system results in large acceptance allowing to record  all three photons from a single event and fast signals from plastic scintillators together with dedicated readout electronics provide a resolution of the time of photon interaction in the detector of about \SI{80}{\pico\second}~\cite{Moskal:2014sra, Moskal:2014rja}, superior to the setups of previously conducted experiments~\cite{Yamazaki:2009hp,vetter:2003}.
%%%%%%%%%%%%%%%%%%%
We expect to significantly improve sensitivity for the CPT test by at least an order of magnitude with respect to the experiment performed using the Gammasphere~\cite{vetter:2003}
by collecting about two orders of magnitude larger statistics due to the possibility of longer runs and  due to the usage of the higher rate of the positronium production
(10~MBq at J-PET vs. 0.4~MBq at Gammasphere) which was limited by pile-ups and 1~$\mu$s coincidence window.
This limitation is overcome by J-PET detector due to its much higher granularity and about two orders of magnitudes shorter duration of signals
leading to the significant reduction of pile-ups and due to the triggerless DAQ~\cite{Korcyl:2014bams} with no hardware coincidence window.
Additional factor is due to angular resolution, which at J-PET is $0.5\degree$ and $\sim1\degree$ for polar and azimuthal angles, respectively, while
at Gammasphere both are about $4\degree$.

With the J-PET detector we expect to improve the sensitivity for measuring expectation values of CP odd operator by more than an order of magnitude with respect to
the measurement performed at Tokyo University~\cite{Yamazaki:2009hp} because of more than two orders of magnitude larger statistics and about 3 times better angular resolution.
In addition, it is possible with J-PET to register any orientation of the decay plane with respect to the spin direction and any relative angle between the gamma quanta, while it was fixed in the previous experiment~\cite{Yamazaki:2009hp}.
Moreover since any of the scintillator strips may detect any of the three gamma quanta, the J-PET is less sensitive to geometric asymmetries of the relative detector arrangement and asymmetries due to the  uncertainties in the detection efficiency determination.

It is also important to stress that the J-PET detector time resolution is a few times higher than at the experiment at Tokyo University~\cite{Yamazaki:2009hp} and by more than an order of magnitude better with respect to the Gammasphere. This will allow us to reduce the background significantly and hence to improve the sensitivity for the studies
of the C violating \mbox{p-Ps$\to3\gamma$} process.

The J-PET system is a multipurpose detector, which allows for positronium polarization determination, as it is described in Section~\ref{sec:advantages}.
Other experiments, like described in reference~\cite{Badertscher:2002nh}, aiming at search of the phenomena beyond the Standard Model
description of \ops/ decays were optimized for the acceptance, efficiency end
energy resolution for registration of gamma quanta
compromising
both timing and angular resolutions.
The latter is one of the most important characteristics in the case of the correlations studies.
The usage of BGO crystals in previous experiments gives and advantage of higher detection efficiency with respect to the organic scintillators used by J-PET,
however this could be compensated by the application of additional layers to J-PET system.
On the other hand, the granularity of the scintillators at J-PET results in the high angular resolution for gamma registration as mentioned before,
while fast timing of the organic scintillators allows for the usage of high activity sources (10~MBq  at J-PET  vs.  $3.6\times10^{-3}$~MBq~\cite{Badertscher:2002nh})
without pile-up contribution.

In the experiments performed so far, the \ops/ decay place was assumed to lie within positronium aerogel targets (e.g. a hemisphere~\cite{vetter:2003}) and no attempt was made to reconstruct the exact point nor the time of its decay. In J-PET, however, due to its relatively high angular acceptance and timing resolution, a reconstruction of the \ops/$\to 3\gamma$ process is possible by means of a new trilateration-based reconstruction method presented in the next Section. Advantages of such full reconstruction of the \ops/ decay for its polarization determination and the CP and CPT symmetry tests with J-PET are discussed in Section~\ref{sec:advantages}.

\section{Principle of the \ops/$\to 3\gamma$ decay reconstruction}
\label{sec:reconstruction}
In the \ops/$\to 3\gamma$ decay, photons travel from the decay point (which needs to be localized) to a particle detector where the places and times of their interaction with the detector are recorded and serve as reference points, further on referred to as photon hits. The lack of a fourth reference is compensated by the fact that all three photons are produced in a three-body decay and thus their momenta as well as the \ops/ decay point are contained within a single plane in the frame of reference of the decaying positronium atom%
\footnote{As the momentum of decaying positronium is not zero, the three photons slightly deviate from coplanarity in the detector frame of reference. This effect was studied using MC simulations and its contribution was found negligible with respect to the $\mathcal{O}$(\SI{1}{\centi\metre}) resolution achieved so far (Section~\ref{sec:simulations}).}.
Hence, an additional constraint needed for the trilateration problem to be determined in this case, can be introduced as a requirement that the decay point lies on a plane spanned by the tree photon recording points.
\begin{figure}[h!]
  \centering
  \subfloat[]{\includegraphics[width=0.40\textwidth]{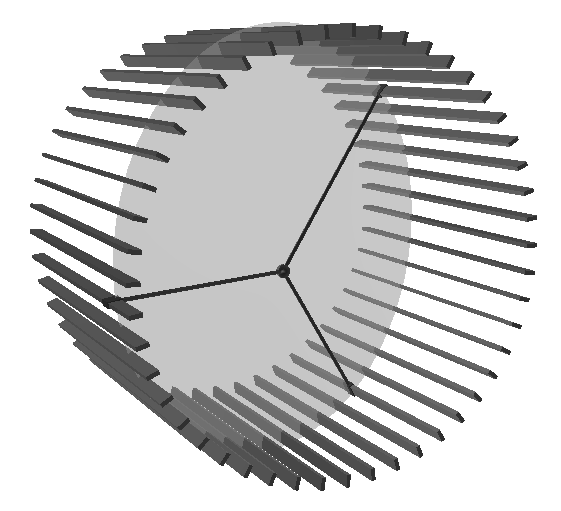}\label{Fig:3d}}
  \subfloat[]{\includegraphics[width=0.45\textwidth]{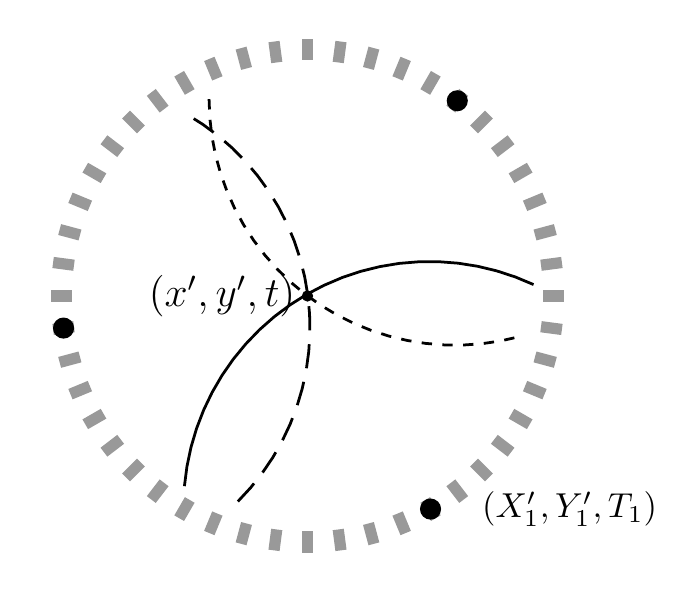}\label{Fig:2d}}
  \caption{
    (a) A scheme of the J-PET detector with photon recording points marked in black. For clarity, only a single layer of the detector is shown and the inter-strip spacing is increased. The gray area indicates the o-Ps$\to 3\gamma$ decay plane. %
    (b) Scheme of the decay reconstruction reduced to a two-dimensional problem in the decay plane. For each recorded photon, its hit coordinates transformed to the decay plane constitute the center of a circle describing possible photon origin points, where radius of the circle depends on the recording time and the unknown \ops/ decay time $t$. The decay point is found as an intersection of the three circles.%
  }
  \label{fig:Nd}
\end{figure}

Figure~\ref{Fig:3d} shows a scheme of an \ops/$\to 3\gamma$ decay taking place inside the J-PET detector. For clarity, only a single layer of the detector is shown and the inter-strip spacing is increased. For each of the photon recording points, its spatial coordinates and time are determined as $(X_i,Y_i,Z_i,T_i),\;i=1,2,3$.
The X and Y coordinates are obtained directly from the known centers of scintillator strip, therefore the accuracy is determined fully by
the size of scintillator, while Z and T are calculated from difference and sum
of signal arrival times at the two ends of scintilator strip, respectively~\cite{Moskal:2013gda}.
Locations of the three points are used to determine their common plane, further on referred to as the decay plane. Next, the spatial coordinates of the hits are transformed from the detector frame of reference to the decay plane. Let $(X'_i,Y'_i,T_i),\;i=1,2,3$ be the transformed coordinates. Hence, the information on coplanarity of the whole event is used to reduce the reconstruction to a two-dimensional problem, presented in Figure~\ref{Fig:2d}.

For a photon hit expressed in the decay plane, its possible origin points constitute a circle centered in the hit point and with a radius equal to the product of the photon's time of flight and the speed of light. However, alike the cases described before, the radii of the three circles are defined up to the time of the photons' creation (measured with respect to the same starting point as the hit recording times) which is unknown. Therefore, if the positronium atom is assumed to decay at time $t$, for the $i$-th hit, the photon origin points $(x',y')$ must satisfy the following equation:
\begin{equation}
  (x'-X'_i)^2 + (y'-Y'_i)^2 = c^2 (T_i - t)^2.
\end{equation}

It is immediately noticed that the \ops/$\to 3\gamma$ decay point, as the common origin of the photons, must lie on an intersection of such circles and thus can be found as a solution of the above equations defined for $i=1,2,3$. Solving this system of equations for $(x',y',t)$ yields the location of the \ops/ decay point as well as time of the decay. In contrast to a classical trilateration problem with constant radii of the circles, at maximum two intersections of the circles defined above can exist.
Therefore, one of the solutions must be discriminated using additional criteria based on the reconstructed decay vertex and time
taking the solution closer to the region of the positronium production target.
The last step of reconstruction is a transformation of the coordinates $(x',y')$ of the chosen solution from the decay plane back to the three-dimensional frame of reference of the detector.

An important property of this reconstruction method is that it allows for obtaining the time of the \ops/$\to 3\gamma$ decay simultaneously with its spatial coordinates, which allows for its application not only to discrete symmetry studies in ortho-positronium decays but also for determination of \ops/ lifetime distribution in the material where it is created, as proposed in a novel concept of medical imaging based on positronium lifetime spectroscopy~\cite{old_patent, new_patent}.
%%% 
\section{Reconstruction performance studies with MC simulations}
\label{sec:simulations}
A Monte Carlo simulation of \ops/$\to 3\gamma$ decays in the J-PET detector was prepared in order to study performance of the reconstruction method presented in Section~\ref{sec:reconstruction}. The simulation included $\beta^{+}$ decays of $^{22}$Na taking place in the center of the device, positrons' propagation in a wall of cylinder-shaped target medium before thermalization, as well as \ops/ creation and its lifetime in the medium. Finally, \ops/$\to 3\gamma$ decays were simulated including dependence of the \ops/$\to 3\gamma$ transition amplitude on the energy of photons according to the predictions based on quantum electrodynamics~\cite{lifshitz_qed} and taking into account non-zero  momentum of the decaying positronium in the detector frame of reference in order to estimate the effect of non-coplanarity of the photons' momenta on reconstruction. Subsequently, the energy dependence of the cross section for Compton scattering in the plastic scintillators was considered in order to simulate the detector response to photons produced in the \ops/ decay. Simulated points and times of photons' interaction in the detector were smeared with the corresponding resolutions of the J-PET detector ($\sigma(t_{hit})$=\SI{80}{\pico\second}, $\sigma(z_{hit})$=\SI{0.93}{\centi\metre}~\cite{Moskal:2014sra}).

The simulated detector consisted of four cylindrical layers of scintillator strips with dimensions of \SI{7x19x500}{\milli\metre} oriented axially, as shown schematically in Fig.~\ref{Fig:3d} for the case of a single layer. An idealized acceptance was assumed, where the inter-strip spaces within layers were maximally reduced. The diameter of the innermost detector layer amounts to \SI{427.8}{\milli\metre} and consists of \num{384} strips~\cite{Kowalski:2015bua}. The size of a single scintillator corresponds to the dimensions of the one used in the real J-PET detector. Photon recording points were resolved up to one strip in the XY plane which resulted in an angular resolution of $\Delta\varphi\approx$~\ang{0.5}. However, resolution of the Z coordinate (along the scintillator strips) and of photon recording time in J-PET are correlated and dependent on the signal reconstruction method used due to the specific features of this detector~\cite{Moskal:2014rja,Raczynski:2014poa,Raczynski:2015zca,Moskal:2015jzy}. Moreover, they are subject to a possible future improvement with an advancement of reconstruction methods applied in J-PET based on multi-SiPM readout~\cite{Moskal_sipm}. Therefore, reconstruction studies were performed both for fixed detector resolutions of $\sigma(t_{hit})$=\SI{80}{\pico\second} and $\sigma(z_{hit})$=\SI{0.93}{\centi\metre} presently achieved with prototypes~\cite{Moskal:2014sra} and for these resolutions varying in the range from \SIrange{0}{160}{\pico\second} and from \SIrange{0}{2}{\centi\metre}, respectively.
\begin{figure}[h!]
  \centering
  \includegraphics[width=1.0\textwidth]{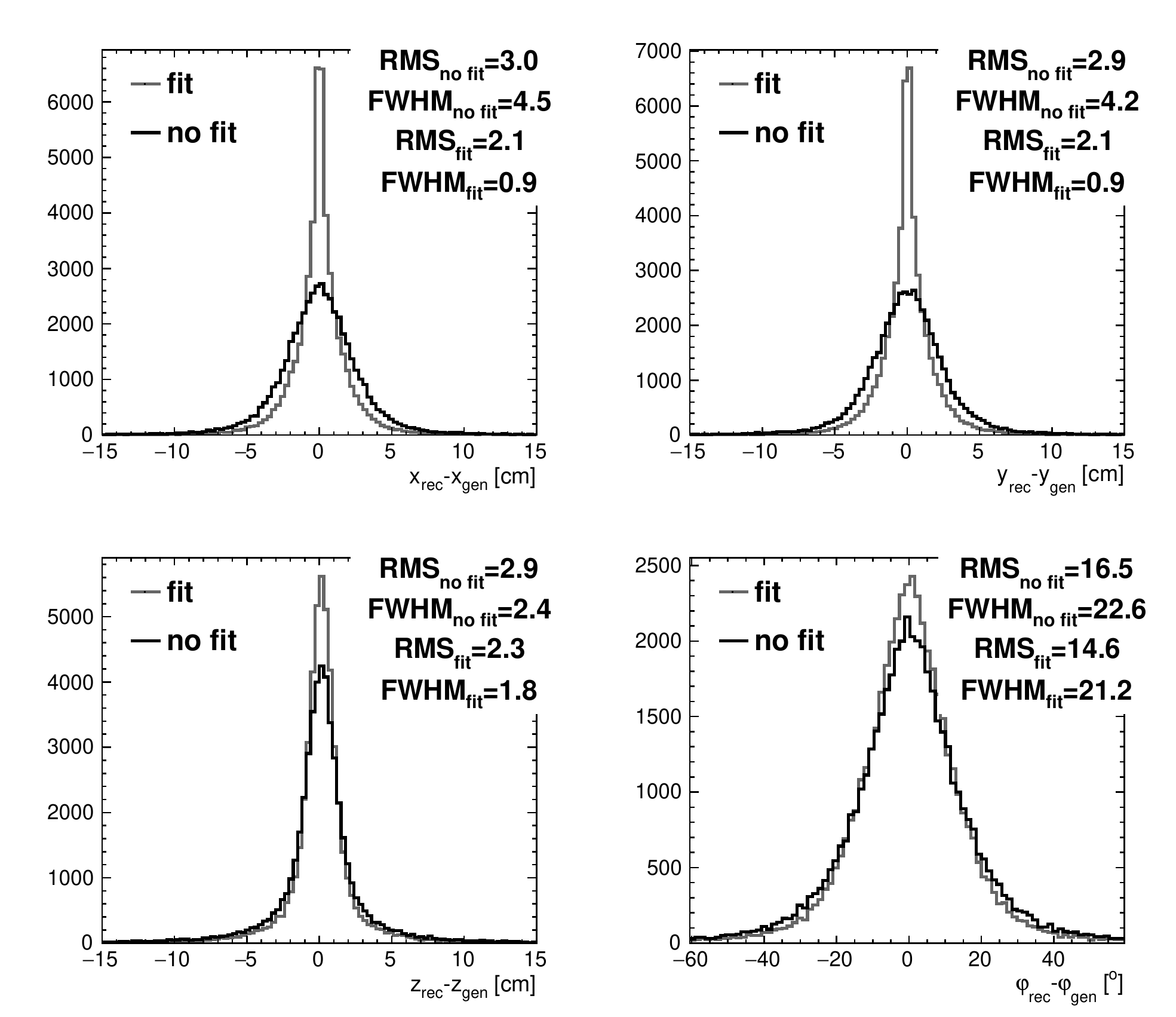}
  \caption{Spatial resolution of the \ops/ decay point in the three Cartesian coordinates and polar angle as obtained with Monte Carlo simulations
(black histogram) with assumed detector resolution $\sigma(t_{hit})$=\SI{80}{\pico\second} and $\sigma(z_{hit})$=\SI{0.93}{\centi\metre}. Solid gray line denotes
results obtained after a kinematic fit of reconstructed events.
The parameters varied within experimental accuracy are the spatial and time coordinates of annihilation gamma quanta interaction points with the detector, 
while the position of the \ops/ decay is constrained to be in the aerogel cylinder wall.}
  \label{fig:3resolutions}
\end{figure}

A sample of \num{e5} events lying fully within detector sensitive area were generated and reconstructed. \ops/ decay points obtained from reconstruction were compared with MC-generated decay places separately in each Cartesian coordinate and the polar angle.
In order to improve performance of the reconstruction a kinematic fit was introduced. For each of the three gamma quanta its time and three spatial coordinates of interaction point in scintilator were varied within their experimental
uncertainty, which for X and Y spatial coordinates corresponds to dimension of the scintilator strip, while for Z and time to detector resolutions ($\sigma(z_{hit})$ and $\sigma(t_{hit})$, respectively).
Since the positronium production medium  is shaped as a thin wall cylinder target, the imposed constraint is equivalent to the vertex being localized in the middle of the cylinder wall in X-Y plane.
Figure~\ref{fig:3resolutions} shows the resulting spectra of reconstruction error when the presently achievable detector resolutions are assumed in simulations.
Application of a kinematic fit yields the RMS of \SI{2.1}{\centi\metre} for X and Y, \SI{2.3}{\centi\metre} for the Z coordinate and \SI{14.6}{\degree} for the polar angle of the reconstructed \ops/ decay point.
The improvement is visible mostly in the resolution of X and Y coordinates due to the applied constraint.
\begin{figure}[h!]
  \centering
  \subfloat[]{\includegraphics[width=0.47\textwidth]{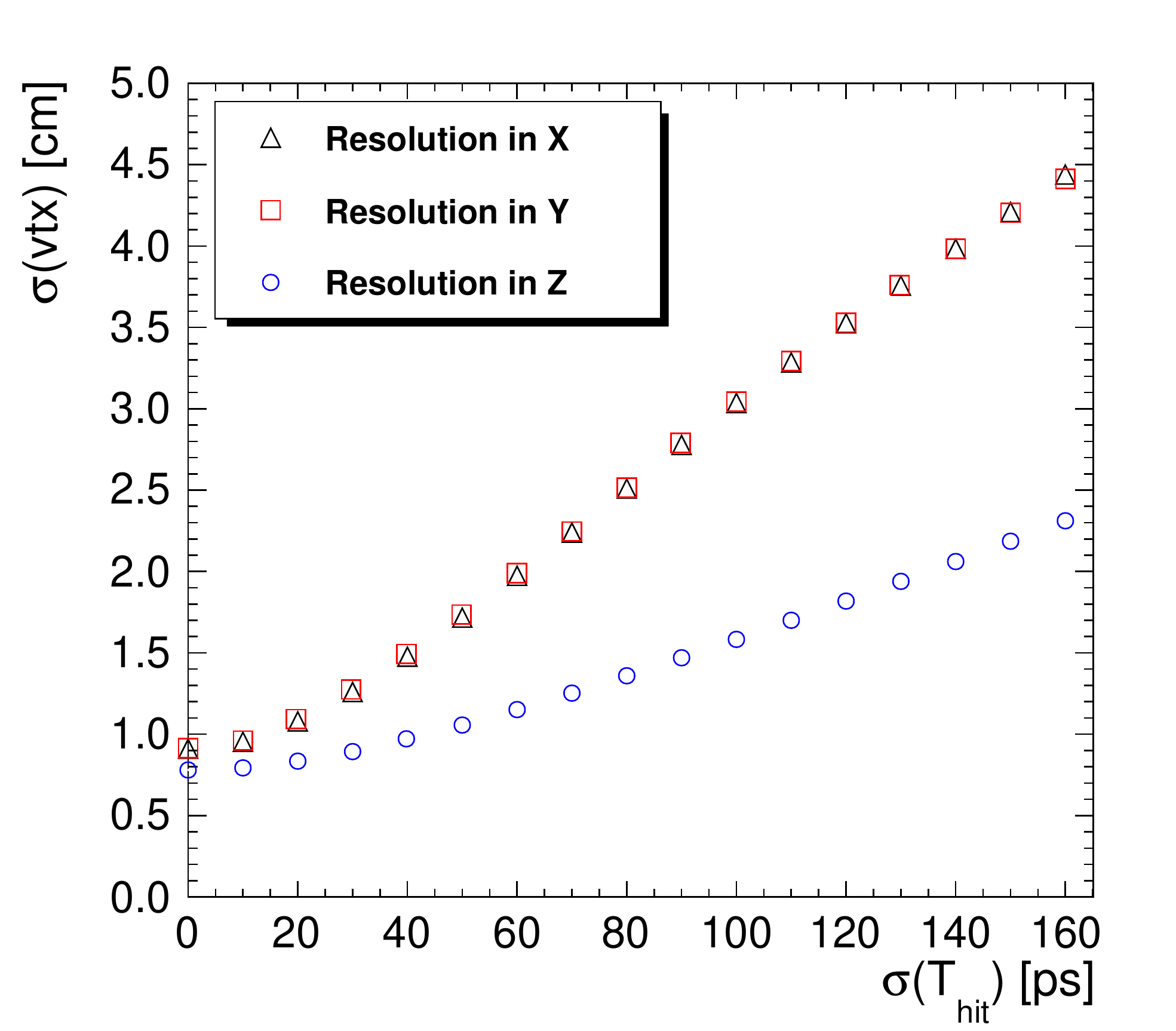}
     \label{Fig:res_vs_t}
  }
  \subfloat[]{
    \includegraphics[width=0.47\textwidth]{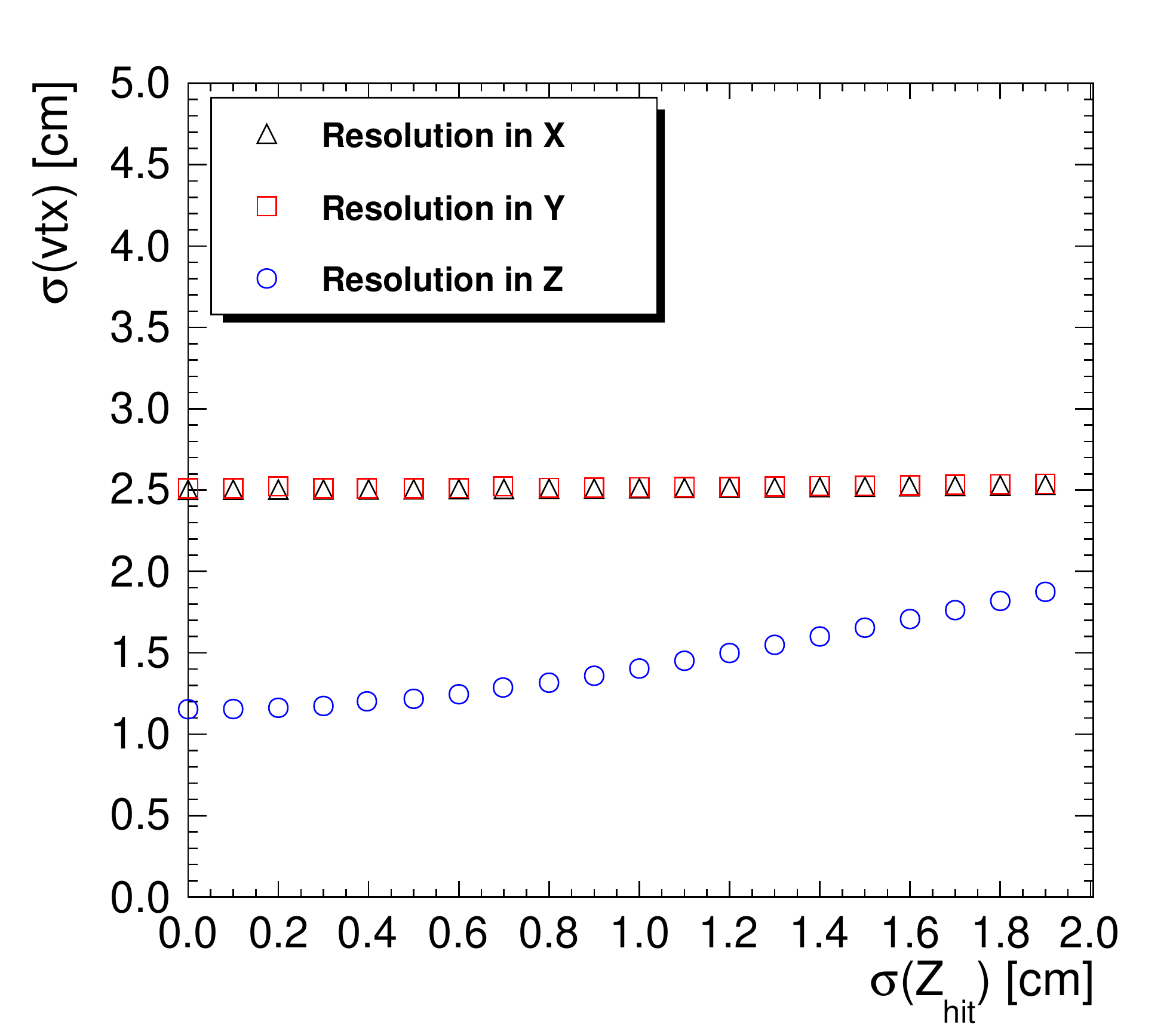}
    \label{Fig:res_vs_z} 
  }
  \caption{Resolution of \ops/ decay point reconstruction in three Cartesian coordinates as a function of detector time resolution with Z resolution of the detector fixed at \SI{0.93}{\centi\metre} (a) and detector resolution in Z with $\sigma(t_{hit})$ fixed at \SI{80}{\pico\second} (b). While the achievable resolution in three dimensions is strongly dependent on the timing capabilities, its dependence on the resolution of the Z~coordinate of a gamma hit is only significant for the Z component of the \ops/ decay place.}
  \label{fig:res_vs}
\end{figure}

Dependence of \ops/$\to 3\gamma$ reconstruction on resolution of photon hits in the detector is presented in Figure~\ref{fig:res_vs}. It is visible that final resolution is mostly dependent on the timing properties of the device rather than on spatial resolution.

\section{Advantages of \ops/$\to 3\gamma$ decay vertex reconstruction}
\label{sec:advantages}
\begin{figure}[h!]
  \centering
  \includegraphics[width=0.5\textwidth]{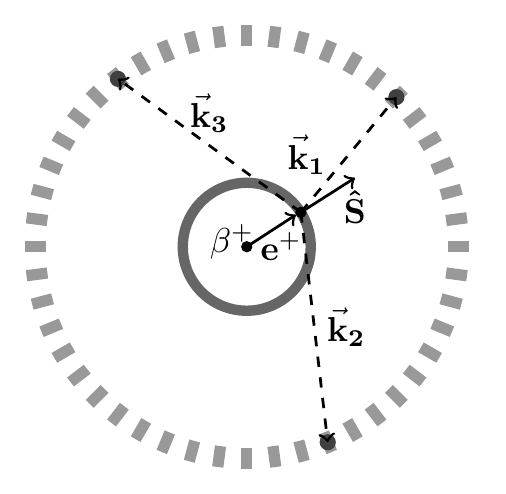}
  \caption{Scheme of a possible \ops/ spin direction determination with the J-PET detector. A $\beta^+$ source is located in the center of a vacuum-filled cylinder covered by aerogel (gray band) in which \ops/ formation and decays take place. Dashed lines denote lines of flight of the three photons used to reconstruct the decay vertex which, in turn, allows to estimate the positron momentum direction and spin direction $\mathbf{\hat{S}}$ of the ortho-positronium.}
  \label{fig:polarization}
\end{figure}

The reconstruction technique presented above offers several advantages for the symmetry tests.
Firstly, it is a simple and efficient way to discriminate random photon coincidences.
%The rates of accidental coincidences as recorded by J-PET have been previously studied and
%for a 10~MBq source and 5~ns coincidence window
%amount to about 20\% of the recorded annihilation rate~\cite{Kowalski:2015bua}.
As shown in Ref.~\cite{Kowalski:2015bua} the two-gamma accidental coincidence rate recorded by the J-PET detector is expected to be about 20\% of the total annihilation rate for a 10~MBq source and 5~ns coincidence window.
While these studies concerned two-gamma coincidences, the rate of accidental triple coincidences for aforementioned source activity is expected to be of the same order
as the average time differences between annihilation events are larger than between recording times of photons from a single event.
Therefore, a typical accidental triple coincidence would be constituted by one annihilation event accompanied by a photon from another annihilation.
The above considerations show that even though good timing properties of the plastic scintillators used in J-PET allow for a very narrow coincidence window,
a fraction of accidental coincidence background at the level of thousands per second irreducible by means of the time window is still expected.
Based on the
dedicated simulations of the setup described in the Ref.~\cite{Kowalski:2015bua} including an aerogel target shaped as a hollow cylinder around a source (as presented in Figure~\ref{fig:polarization}) with an activity of 10~MBq
we studied the accidental triple coincidences and obtained a recorded rate of 2800/s within a time window of 5~ns.
These can be efficiently removed by means of the presented vertex reconstruction.
As the reconstruction technique is especially sensitive to the timing information, events where the three photons used were not emitted at the same time (significantly beyond the \SI{80}{\pico\second} resolution) will be immediately recognized by the reconstructed point and time being either non-physical or significantly shifted off the cylindrical target.
Based on the aforementioned simulations this approach allowed to reject 89\% of the remaining accidental coincidences.
Therefore combination of both the narrow coincidence window and trilateration based reconstruction allows for a rejection of accidentals at a level of 98\%.
%%%%%%%

Secondly, \ops/ decay point reconstruction offers new possibilities to determine the polarization of the decaying positronium. As the ortho-positronia must be polarized in order to study the angular correlations in their decays, previous experiments have either used an external magnetic field~\cite{Yamazaki:2009hp} or limited the linear polarization of positrons used to form \ops/ by geometrically restricting their momenta to a single hemisphere~\cite{vetter:2003}. In the latter approach, one takes advantage of the parity violation in the $\beta^{+}$ decays which results in the polarisation of emitted positrons along their direction of motion with a degree of P=$\frac{\upsilon}{c}$ where $\upsilon$ is the velocity of positron. This polarization is to large extent preserved during the thermalization process and transferred to the positronium~\cite{VanHouse:1984zz,Zitzewitz:1979}. However, allowing for positrons' momenta in a whole hemisphere around the given direction results in a further decrease of polarization by a factor \num{1/2}. 

The ability to reconstruct the \ops/ decay vertex allows for a different way of polarization estimation, presented schematically in Figure~\ref{fig:polarization}~\cite{PawelActaB}. If a $\beta^+$ source is located inside a vacuum-filled cylinder covered with a layer of a medium for positronium formation (such as aerogel), then reconstructed position of the \ops/$\to 3\gamma$ decay can be used to obtain the positron momentum direction.
As the average polarization of positron from a $\beta$ decay along their momentum is \SI{67}{\percent} for $^{22}$Na and \SI{90}{\percent} for $^{68}$Ge as estimated in~\cite{vetter:2003}, the $e^{+}$ momentum direction reconstructed an an event-by-event basis can be further used to statistically estimate the \ops/ polarization.
The average polarization {\em P} of positrons can be expressed as $P=\frac{\upsilon}{c}(1+\texttt{cos}\alpha)/2$ if positrons are emitted
in a cone with an opening angle of $2\alpha$~\cite{coleman_positrons}. 
With the J-PET detector and the cylindrical positronium target with a radius of 10~cm (as presented schematically in Fig.~\ref{fig:polarization}) the uncertainty of determination of positron direction will amount to about 15$\degree$ as shown in Fig.~\ref{fig:3resolutions}.
Therefore the polarization is reduced by 2\% only, while in case of Gammasphere~\cite{vetter:2003}, where positrons were emitted in a whole hemisphere ($\alpha=90\degree$), polarization was reduced by factor of 2.
\section{Summary}
Decays of ortho-positronium atoms into three photons can be reconstructed in terms of time and place of the decay using the trilateration-based method presented in this article. Places and times of recording the three produced gamma quanta in the J-PET detector constitute sufficient input information to obtain an analytical solution for the decay thanks to the specific geometry of the three-body \ops/$\to 3\gamma$ decay where the photon recording points are co-planar with the decay place. Performance of the reconstruction for the novel J-PET device was estimated with Monte Carlo simulations and
the spatial resolution achievable with a kinematic fit is at the level of \SI{2}{\centi\metre} (FWHM) for X and Y and at the level of \SI{1}{\centi\metre} (FWHM) for Z
with the present timing capabilities of J-PET.
The presented reconstruction method can be used in the tests of the CP and CPT symmetries with the J-PET detector, allowing for background rejection as well as determination of \ops/ polarization.

\section*{Acknowledgments}
We acknowledge technical and administrative support of A.~Kubica-Misztal, A.~Heczko, 
M.~Kajetanowicz, G.~Konopka-Cupia\l, W.~Migda\l, and the financial
support by The Polish National Centre for Research and Development through
grant INNOTECH-K1/IN1/64/159174/NCBR/12, the Foundation for  Polish Science
through MPD programme, the EU and MSHE Grant  No.  POIG.02.03.00-161 00-
013/09, Doctus - the Lesser Poland  PhD Scholarship  Fund, Marian
Smoluchowski Krakow Research Consortium ”Matter-Energy-Future" and by 
The Polish National Centre for Research and Development through LIDER program grant no. 274/L-6/2014.

\section*{References}
\bibliography{jpet_ops_rec}
\end{document}